\documentclass[10pt,reqno]{amsart}
\usepackage{amssymb,mathrsfs,color}
\usepackage{pinlabel}

\usepackage{amsmath, amsfonts, amsthm, verbatim, amssymb}
\usepackage{epstopdf, mathtools}
\usepackage{color}
\usepackage{esint}
\usepackage{stmaryrd}
\usepackage{graphicx}
\usepackage{xcolor} % A package to add color.
\usepackage{tensor}
\usepackage{slashed}
\usepackage{cite}
\mathtoolsset{showonlyrefs}
\usepackage{caption, subcaption}

\newcommand{\bs}[1]{\boldsymbol{#1}}

\newcommand{\la}{\lambda}

\newcommand{\p}{\partial}

\usepackage[multiple]{footmisc}
\numberwithin{equation}{section}

\theoremstyle{remark}

\renewcommand{\div}{\mathrm{div}\,}
\newcommand{\rar}{\rightarrow}
\newcommand{\bbG}{\mathbb G}

\newcommand{\tens}{\otimes}

\begin{document}

\title[Second-Gradient Inclined Fluid Flows]{Inclined flow of a second-gradient incompressible fluid with pressure-dependent viscosity} 
\author{C. Balitactac and C. Rodriguez}

\begin{abstract}
Many viscous liquids behave effectively as incompressible under high pressures but display a pronounced dependence of viscosity on pressure. The classical incompressible Navier–Stokes model cannot account for both features, and a simple pressure-dependent modification introduces questions about the well-posedness of the resulting equations. This paper presents the first study of a second-gradient extension of the incompressible Navier–Stokes model, recently introduced by the authors, which includes higher-order spatial derivatives, pressure-sensitive viscosities, and complementary boundary conditions. Focusing on steady flow down an inclined plane, we adopt Barus’ exponential law and impose weak adherence at the lower boundary and a prescribed ambient pressure at the free surface. Through numerical simulations, we examine how the flow profile varies with the angle of inclination, ambient pressure, viscosity sensitivity to pressure, and internal length scale.
\end{abstract}

\maketitle

\section{Introduction}

Bridgman’s Nobel Prize–winning investigations revealed that many viscous liquids, when subjected to high pressures, behave essentially as incompressible while simultaneously showing a strong pressure dependence on their viscosity \cite{Bridgman1926, Bridgman1931}. This dual behavior, apparent incompressibility coupled with pressure-sensitive viscous response, is not reflected in the classical incompressible Navier–Stokes equations, which assume a constant viscosity independent of pressure.

To better capture such high-pressure phenomena, a number of studies have considered a modified version of the incompressible Navier–Stokes model, formulated within the framework of classical continuum mechanics \cite{Bair1998, Rajagopal2003, Bayada2013, Gustafsson2015, Denn1981, Hron2001, Vasudevaiah2005, Prusa2010, Kalogirou2011, Rajagopal2012, Rajagopal2013, Marusic-Paloka2013, Housiadas2015, PrusaRehor2016, Gwynllyw1996, Hron2003, Chung2010, Knauf2013, Janecka2014}. In this model, viscosity is treated as a function of pressure, while the incompressibility constraint is retained:
\begin{align}
	\rho \dot{\boldsymbol{v}} = \div \boldsymbol{T} + \rho \bs b, \quad \boldsymbol{T} = -p\boldsymbol{I} + 2\hat{\mu}(p) \boldsymbol{D}, \quad \div \bs v = 0. \label{eq:pressuredependent}
\end{align}
Here, $\rho$ is the constant mass density, $\bs v$ is the velocity field, $\dot{\mbox{}}$ denotes the material time derivative, $p$ is the pressure, $\hat \mu(p) > 0$ is the pressure-dependent viscosity, $\bs D = \frac{1}{2}(\nabla \bs v + \nabla \bs v^T)$ is the stretching, and $\bs b$ is a body force per unit mass. A widely used empirical relation for $\hat \mu(p)$ is Barus' exponential law \cite{Barus1893}:
\begin{align}
	\hat \mu(p) = \mu_0 \exp(\beta p), \label{eq:barus}
\end{align}
where $\mu_0 > 0$ and $\beta > 0$ characterizes the sensitivity of viscosity to changes in pressure.

Although this pressure-dependent formulation introduces important physical effects absent in the classical theory, it also gives rise to significant mathematical complications. In particular, the equation determining the pressure may lose ellipticity, rendering the system undetermined or ill-posed; see the discussion in Section 3 of \cite{BalitactacRodriguez25}. Local well-posedness has only been established under the condition $\hat{\mu}(p)/p \to 0$ as $p \to \infty$, a requirement incompatible with Barus' law and experimental observations \cite{Renardy1986}. Additional results are largely restricted to steady, low-speed flows in bounded domains \cite{GazzolaSecchi1998}.

In recent work \cite{BalitactacRodriguez25}, we proposed a \textit{second-gradient} incompressible viscous fluid model that retains the essential pressure-dependent viscosity. By augmenting the classical theory with second-gradient contributions, the resulting pressure equation remains elliptic regardless of the velocity field, avoiding the determinability issues inherent to \eqref{eq:pressuredependent}; see Section 3 of \cite{BalitactacRodriguez25}. The model introduced in \cite{BalitactacRodriguez25} is characterized by the standard Cauchy stress $\bs T$ and a third-order \textit{hyperstress} tensor $\bbG$ given by:
\begin{align}
	\begin{split}
		\bs T &= -p \bs I + 2\hat \mu(p) \bs D, \\
		\bbG &= -\frac{\ell_1^2}{2}\Bigl [\bs I \tens \nabla p + (\bs I \tens \nabla p)^{T(2,3)} \Bigr ] + \mu_0(\ell_2^2 + \ell_3^2) \nabla \nabla \bs v \\&\quad+ \frac{1}{2}\mu_0(\ell_2^2 -\ell_3^2) \Bigl [
		\nabla \nabla \bs v^{T(1,2)} + \nabla \nabla \bs v^{T(1,3)} - \Delta \bs v \tens \bs I  
		\Bigr ] \\
		&\quad +  \mu_0\Bigl (-\frac{1}{16}\ell_2^2 + \frac{1}{4} \ell_3^2 - \frac{1}{2} \ell_4^2 \Bigr )\Bigl [
		\bs I \tens \Delta \bs v + (\bs I \tens \Delta \bs v)^{T(2,3)} - 4 \Delta \bs v \tens \bs I 
		\Bigr ],
	\end{split} \label{eq:Gequations}
\end{align}
with the internal length scales $\ell_1, \dots, \ell_4$ related via
\begin{align}
	\ell_1^2 = \frac{3}{4}\ell_2^2 + \frac{1}{2}\ell_3^2 + 2\ell_4^2.
\end{align}
By applying the principle of virtual power \cite{Germain73a, Germain73b, FriedGurtin06}, the governing equations reduce to the incompressibility condition $\div \bs v = 0$ and
\begin{equation}\label{eq:Gpressuredependent}
	\rho \left [\dot{\bs v} - \ell_0^2 \div(\dot{\overline{\nabla \bs v}}) \right ] = -\nabla(p - \ell_1^2 \Delta p) + \div (2\hat \mu(p) \bs D) - \ell_1^2 \mu_0 \Delta^2 \bs v + \rho \bs b,
\end{equation}
where $\ell_0$ is an additional internal inertial length scale.

While \cite{BalitactacRodriguez25} focused on cylindrical flows with constant viscosity, this paper presents the first analysis of steady flow for the full second-gradient model with pressure-dependent viscosity, examining gravity-driven motion down an inclined plane of angle $\alpha$ and depth $h$ (see Figure~\ref{fig:f1}). The fluid adheres weakly to the lower boundary, while the upper surface is exposed to a prescribed ambient pressure; see \eqref{eq:bdrycnd}. We adopt Barus’ exponential law \eqref{eq:barus} and establish the existence and uniqueness of solutions to the resulting boundary value problem. In Section 3, we then numerically explore how the flow profile is influenced by variations in the angle of inclination, and the nondimensionalized ambient pressure $p_1$, the Barus parameter $\beta$, and the internal length scale $\ell_1$.

\subsection*{Acknowledgments} C. B. and C. R. gratefully acknowledge support from NSF DMS-2307562, and C.B. also appreciates support from NSF RTG DMS-2135998.

\section{Planar inclined Flow}

\begin{figure}[t]
	\centering
	\includegraphics[width=.6\linewidth]{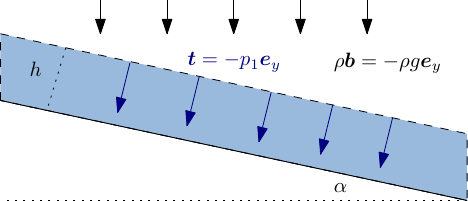}
	\caption{The set-up of inclined flow.}
	\label{fig:f1}
\end{figure}

In this work we examine the steady flow of a second-gradient fluid with pressure-dependent viscosity of depth $h$ under the force of gravity down an inclined plane making an angle $\alpha$ with the horizontal (see Figure \ref{fig:f1}). We rotate the axes so that the inclined plane corresponds to the $xz$-plane, and the body force is then given by
\begin{equation}
	\rho \bs b = \rho g (\sin \alpha\, \bs e_x - \cos \alpha \,\bs e_y).
\end{equation}
The field equations \eqref{eq:Gpressuredependent} are complemented by the boundary conditions expressing \textit{weak adherence} along the plane $y = 0$ and constant pressure $p_1$ (e.g., atmospheric pressure) at $y = h$: 
\begin{equation}\label{eq:vbdrycnd}
	\begin{cases}
		\bs v = \bs 0, \quad \bs m = \bs 0, \quad \frac{\p p}{\p \bs n} = 0, & \mbox{at }y = 0 \\
		\bs t = -p_1\bs e_y, \quad \bs m = \bs 0, \quad \frac{\p p}{\p \bs n} = 0, & \mbox{at }y = h.
	\end{cases}
\end{equation}
Here $\bs t$ and $\bs m$ are the traction and hypertraction (see \cite{FriedGurtin06, BalitactacRodriguez25}), given, respectively, by 
\begin{gather}
\bs t = \bs T \bs n - (\div \bbG)\bs n - \div_s(\bbG \bs n) - 2K \bbG[\bs n \otimes \bs n], \quad \bs m = \bbG[\bs n \tens \bs n], 
\end{gather}
where $\bs T$ and $\bbG$ are described in \eqref{eq:Gequations}, $\div_s$ denotes the surface divergence and $K$ is the mean curvature. 

We assume the shear flow ansatz
\begin{gather}
	\bs v = v(y)\bs e_x, \quad p = p(y), \quad y \in [0,h],
\end{gather}
and thus, 
\begin{equation}
	%	2 \mu [\bs D] = \begin{bmatrix}
		%		0 & \mu v' & 0 \\
		%		\mu v' & 0 & 0 \\
		%		0 & 0 & 0
		%	\end{bmatrix} \text{ and }
	[\div 2 \mu \bs D] = \begin{bmatrix}
		(\mu v')' \\ 0 \\ 0
	\end{bmatrix},
\end{equation}
where $\mu = \hat \mu(p) = \mu_0 \exp(\beta p)$.
The governing field equations \eqref{eq:Gpressuredependent} reduce to the following ordinary differential equations with $\ell := \ell_1$:
\begin{equation}\label{eq:fieldeqns}
	\begin{cases}
		0 = (\mu v')' - \mu_0\ell^2v'''' + \rho g \sin\alpha \\
		0 = -(p' - \ell^2 p''') - \rho g \cos\alpha.
	\end{cases}
\end{equation}
The boundary conditions \eqref{eq:vbdrycnd} are equivalent to
\begin{equation}\label{eq:bdrycnd}
	\begin{cases}
		v(0) = 0, v''(0) = 0, p'(0) = 0 \\
		v''(h) = 0, p'(h) = 0 \\
		\mu v'(h) - \mu_0 \ell^2 v'''(h) = 0,
        \\\ell^2 p''(h) - p(h) = -p_1 
	\end{cases}.
\end{equation}

Integrating the second equation in \eqref{eq:fieldeqns} from $y$ to $h$ and using the boundary conditions \eqref{eq:bdrycnd} gives the following:
\begin{align}
	0 &= - p(h) + \ell^2 p''(h) - \rho g h \cos\alpha + p(y) - \ell^2 p''(y) + \rho g y \cos\alpha \\
	&= -p_1 -\rho g h \cos \alpha + p(y) - \ell^2 p''(y) + \rho g y \cos\alpha.
\end{align}
Thus, the equation that the pressure satisfies is
\begin{equation}\label{eq:pressureeqn}
\begin{cases}
	\ell^2 p''(y) - p(y) = -p_1 - (h-y)\rho g \cos\alpha, \quad y \in [0,h], \\
	p'(0) = p'(h) = 0.
\end{cases}
\end{equation}
%Since this is a linear ordinary differential equation with constant coefficients, the solution $p$ has the following form:
%\begin{equation}
%	p(y) = c_1 e^{y/\ell} + c_2 e^{-y/\ell} + p_1 + (h-y)\rho g \cos\alpha.
%\end{equation}
One readily finds that $p$ has the specific form
\begin{equation}
	p(y) = \frac{\ell \rho g\cos\alpha}{2\sinh(h/\ell)}\Bigl((1-e^{-h/\ell})e^{y/\ell} + (1- e^{h/\ell})e^{-y/\ell}\Bigr) + p_1 + \rho g (h-y)\cos\alpha.
\end{equation}
In the classical continuum setting with $\ell = 0$, we have 
\begin{align}
	p_c(y) = p_1 + \rho g (h-y) \cos \alpha. \label{eq:ellequals0pressure}
\end{align}

Integrating the first equation of \eqref{eq:fieldeqns} from $y$ to $h$ and using the boundary conditions yields:  
\begin{align}
	0 &= \mu(p(h)) v'(h) - \mu_0 \ell^2 v'''(h) + \rho g h\sin\alpha - \mu(p(y)) v'(y) 
	\\&\quad + \mu_0\ell^2v'''(y) - \rho g y \sin\alpha \\
	&= \rho g h\sin\alpha - \mu(p(y)) v'(y) + \mu_0\ell^2v'''(y) - \rho g y \sin\alpha.
\end{align}
Thus, the velocity is governed by the boundary value problem
\begin{equation}\label{eq:velocitypressuredep}
	\begin{cases}
		\mu_0 \ell^2 v'''(y) - \mu(p(y)) v'(y) = -\rho g (h-y)\sin\alpha, \quad y \in [0,h], \\
		v(0) = 0, v''(0) = v''(h) = 0.
	\end{cases}
\end{equation}
If $\beta = 0$, we obtain the following solution $v_{\beta = 0}(\sigma)$ to \eqref{eq:velocitypressuredep} corresponding to pressure-independent viscosity: 
\begin{equation}\label{eq:constviscsoln}
    v_{\beta = 0}(y) = \frac{g\rho\sin\alpha}{2\mu_0}\Bigl(2hy - y^2 - 2\ell^2 + 2\ell^2 \cosh\Bigl(\frac{h-2y}{2\ell}\Bigr)\mathrm{sech}\Bigl(\frac{h}{2\ell}\Bigr) \Bigr)
\end{equation}
The pressure-dependent viscosity model rooted in classical continuum mechanics corresponds to 
\begin{equation}\label{eq:classicalpressuredependent}
	\begin{cases}
		- \mu(p(y)) v'(y) = -\rho g (h-y)\sin\alpha, \quad y \in [0,h], \\
		v(0) = 0,
	\end{cases}
\end{equation}
with $p(y)$ given by \eqref{eq:ellequals0pressure}. The solution to \eqref{eq:classicalpressuredependent}, $v_c(y)$, is readily found to be
\begin{align}\label{eq:classicalpressuredependentsoln}
	v_c(y)& = \frac{\sec\alpha\tan\alpha \exp(-\beta(gh\rho\cos\alpha + p_1))}{g\beta^2 \rho\mu_0} \\
	&\times \biggl[ \exp(yg\beta\rho\cos\alpha)(1+g\beta\rho\cos\alpha(h-y)) - gh\beta\rho\cos\alpha - 1\biggr]
\end{align}
with the classical solution to \eqref{eq:classicalpressuredependent} corresponding to pressure-independent viscosity ($\beta = 0$) given by 
\begin{align}\label{eq:classicalsoln}
v_{c, \beta = 0}(y) = \frac{g \rho \sin \alpha}{2\mu_0}(2hy - y^2). 
\end{align}

We now non-dimensionalize the variables. We define
\begin{gather}
	\sigma = \frac{y}{h}, \quad \lambda = \frac{\ell}{h}, \quad u(\sigma) = \frac{\mu_0}{\rho g h^2} v(h \sigma), 
	\\\pi(\sigma) = \frac{1}{\rho g h} p(h\sigma), \quad \pi_1 = \frac{p_1}{\rho g h}, \quad \gamma = \beta \rho g h. 
\end{gather}
The nondimensionalized pressure is given by
\begin{equation}
	\pi(\sigma) = \frac{\lambda \cos\alpha}{2\sinh(1/\lambda)}\Bigl((1-e^{-1/\lambda})e^{\sigma/\lambda} + (1- e^{1/\lambda})e^{-\sigma/\lambda}\Bigr) + \pi_1 + (1-\sigma)\cos\alpha,
\end{equation}
the dimensionless velocity profile $u_c$ corresponding to $v_c$ is given by
\begin{align*}
	u_c(\sigma) &= \frac{\sec\alpha\tan\alpha \exp(-\gamma(\cos\alpha + \pi_1))}{\gamma^2} \\
	&\times \biggl[\exp(\sigma\gamma\cos\alpha)(1+(1-\sigma)\gamma\cos\alpha) - \gamma\cos\alpha - 1\biggr],
\end{align*}
and the dimensionless velocity profile $u_{\gamma = 0}$ corresponding to $v_{\beta = 0}$ is given by
\begin{equation}
    u_{\gamma = 0}(\sigma) = \sin\alpha \Bigl( \sigma - \frac{\sigma^2}{2} - \lambda^2 \Bigl(1 - \cosh\Bigl(\frac{1-2\sigma}{2\lambda}\Bigr)\mathrm{sech}\Bigl(\frac{1}{2\lambda}\Bigr)\Bigr) \Bigr ).
\end{equation}
The dimensionless pressure $\pi_c$ corresponding to $p_c$ and velocity profile $u_{c,\gamma=0}$ corresponding to $v_{c,\beta = 0}$ are given by 
\begin{gather}
    \pi_c(\sigma) = \pi_1 + (1-\sigma) \cos \alpha, \quad  
    u_{c,\gamma=0}(\sigma) = \sin \alpha \Bigl ( \sigma - \frac{\sigma^2}{2} \Bigr ). 
\end{gather}
We note that straightforward arguments show that $\pi(\sigma) \to \pi_c(\sigma)$ and $u_{\gamma = 0} \rar u_{c, \gamma =0}$ as $\lambda \to 0$ pointwise. Indeed, it is sufficient to show the following pointwise limits:
\begin{gather}
     \frac{\lambda \cos\alpha}{2\sinh(1/\lambda)}\Bigl((1-e^{-1/\lambda})e^{\sigma/\lambda} + (1- e^{1/\lambda})e^{-\sigma/\lambda}\Bigr) \to 0 \quad \text{ as } \lambda \to 0, \label{eq:lim1} \\
    \la^2 \cosh\Bigl (\frac{1-2\sigma}{2\lambda} \Bigr ) \mathrm{sech} \Bigl ( \frac{1}{2\lambda} \Bigr ) \to 0 \quad \text{ as } \lambda \to 0.  \label{eq:lim2}
\end{gather}
Writing $\sinh(1/\lambda)$ in its exponential form and multiplying the left-hand side of \eqref{eq:lim1} by $e^{-1/\lambda}/e^{-1/\lambda}$ and letting $\lambda \to 0$ yields the limit \eqref{eq:lim1}:
\begin{align}
    &\Bigl(\frac{\lambda \cos\alpha(e^{\sigma/\lambda} - e^{(\sigma - 1)/\lambda} + e^{-\sigma/\lambda} - e^{(1-\sigma)/\lambda)}}{e^{1/\lambda} - e^{-1/\lambda}}\Bigr)\frac{e^{-1/\lambda}}{e^{-1/\lambda}} \\
    &= \lambda\cos\alpha\Bigl( \frac{e^{(\sigma - 1)/\lambda} - e^{(\sigma - 2)/\lambda} + e^{-(\sigma+1)/\lambda} - e^{-\sigma/\lambda}}{1 - e^{-2/\lambda}}\Bigr) \to 0 \quad \text{ as } \lambda \to 0.
\end{align}
A similar argument using that $1-2\sigma \in [-1,1]$ proves \eqref{eq:lim2}.

In terms of the dimensionless variables, the governing equation \eqref{eq:velocitypressuredep} is equivalent to 
\begin{equation}\label{eq:nondvelocitypressuredep}
	\begin{cases}
		\la^2 u'''(\sigma) - \exp(\gamma \pi(\sigma)) u'(\sigma) = -(1-\sigma)\sin\alpha, \quad \sigma \in [0,1], \\
		u(0) = 0, u''(0) = u''(1) = 0,
	\end{cases}
\end{equation}
with 
\begin{equation}
	\pi(\sigma) = \frac{\lambda \cos\alpha}{2\sinh(1/\lambda)}\Bigl((1-e^{-1/\lambda})e^{\sigma/\lambda} + (1- e^{1/\lambda})e^{-\sigma/\lambda}\Bigr) + \pi_1 + (1-\sigma)\cos\alpha.
\end{equation}
We conclude this section by showing that \eqref{eq:nondvelocitypressuredep} always has a unique solution. We set $f = u'$. Then $f$ must satisfy 
\begin{equation}\label{eq:nondvelocitypressuredep2}
	\begin{cases}
		\la^2 f''(\sigma) - \exp(\gamma \pi(\sigma)) f(\sigma) = -(1-\sigma)\sin\alpha, \quad \sigma \in [0,1], \\
		f'(0) = f'(1) = 0,
	\end{cases}
\end{equation}
The homogeneous equation
\begin{align}
\begin{cases}
-\la^2 g''(\sigma) + \exp(\gamma \pi(\sigma))g(\sigma) = 0, \label{eq:homeq} \\
g'(0) = g'(1) = 0,
\end{cases}
\end{align}
only has the trivial solution $g = 0$ since multiplication of $\eqref{eq:homeq}_1$ by $g$ and integration by parts implies 
\begin{align}
    \int_0^1 \la^2 [g'(\sigma)]^2 + \exp(\gamma \pi(\sigma))[g(\sigma)]^2\, d\sigma = 0. 
\end{align}
Since $\exp(\gamma \pi(\sigma))$ is bounded from below by a positive constant on $[0,1]$, we conclude that $\int_0^1 [g(\sigma)]^2d\sigma = 0$, and thus, $g = 0$. 
In summary, \eqref{eq:nondvelocitypressuredep2} can be solved uniquely for each $\lambda$, $\gamma$, and $\alpha$ using the method of variation of parameters. The dimensionless velocity is then obtained uniquely from the no-slip condition $u(0) = 0$ via $u(\sigma) = \int_0^\sigma f(s)ds$. 

\section{Numerical solutions}
\begin{figure}[b]
    \centering
    \begin{subfigure}[b]{0.49\textwidth}
        \centering
        \includegraphics[trim = {0 6cm 0 6cm},clip,width=\textwidth]{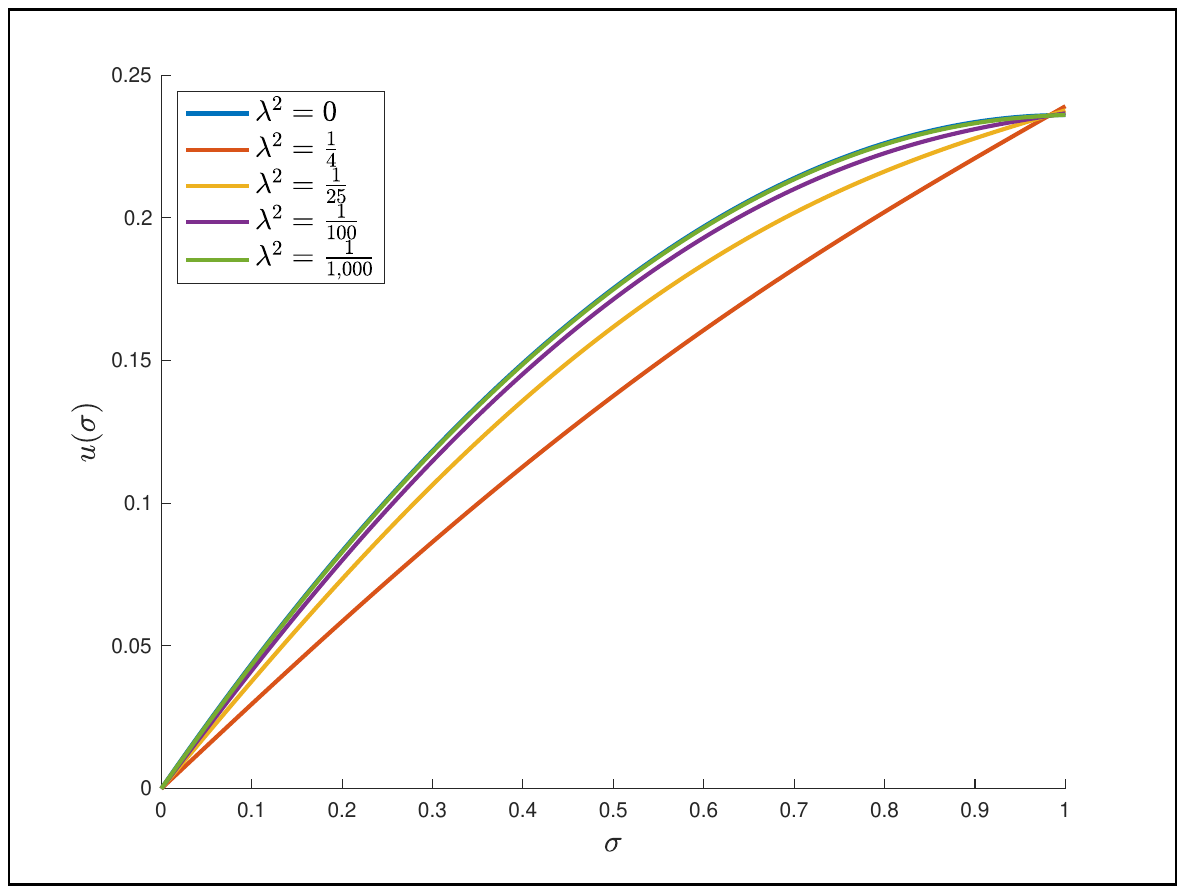}
        \caption{$\alpha = \pi/6, \pi_1 = 0, \gamma = 1/10$}
        \label{fig:f2}
    \end{subfigure}
    \hfill
    \begin{subfigure}[b]{0.49\textwidth}
        \centering
        \includegraphics[trim = {0 6cm 0 6cm},clip,width=\textwidth]{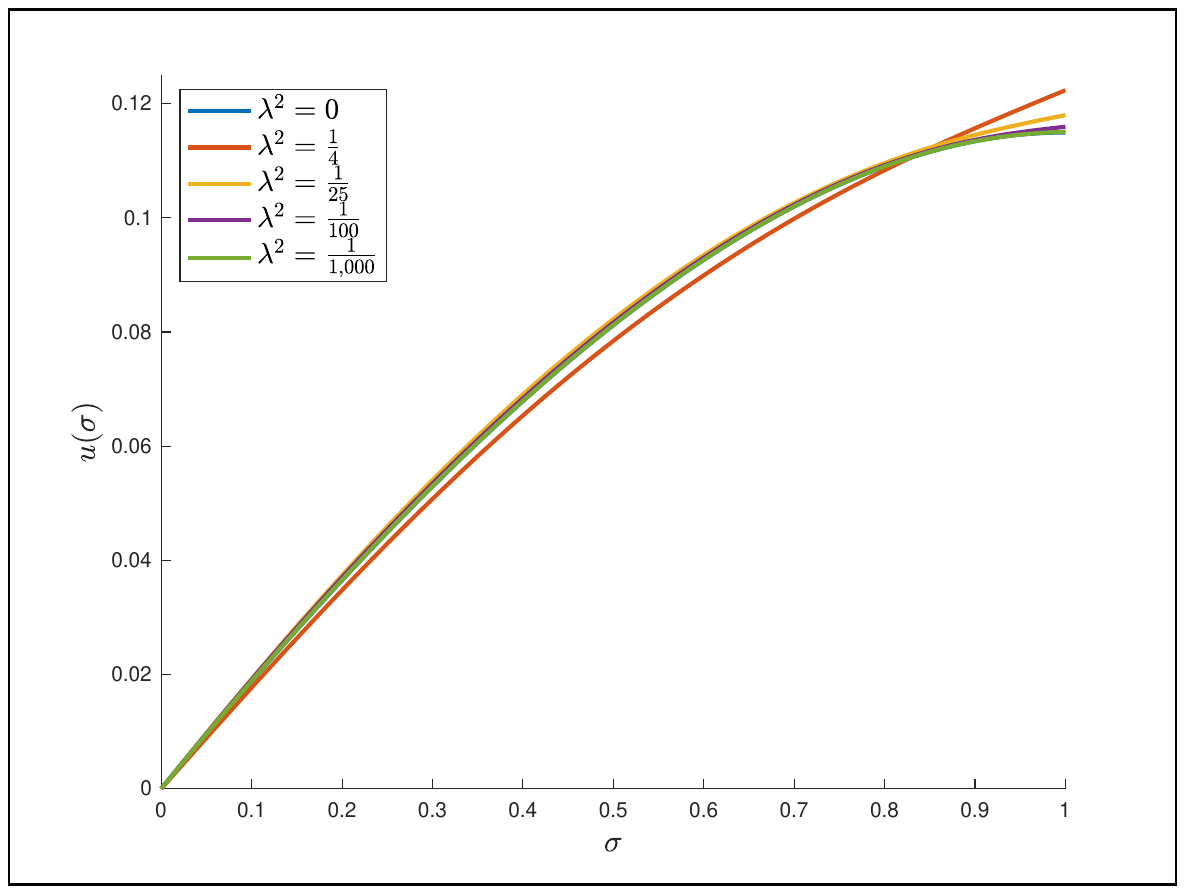}
        \caption{$\alpha = \pi/3, \pi_1 = 1, \gamma = 1$}
        \label{fig:f3}
    \end{subfigure}
    \caption{Graphs of the dimensionless velocity $u(\sigma)$ with varying values of $\lambda^2$. The curve corresponding to $\lambda^2 = 0$ represents the classical solution $u_c(\sigma)$.}
\end{figure}

\begin{figure}
    \centering
    \begin{subfigure}[b]{0.49\textwidth}
        \centering
        \includegraphics[trim = {0 6cm 0 6cm},clip,width=\textwidth]{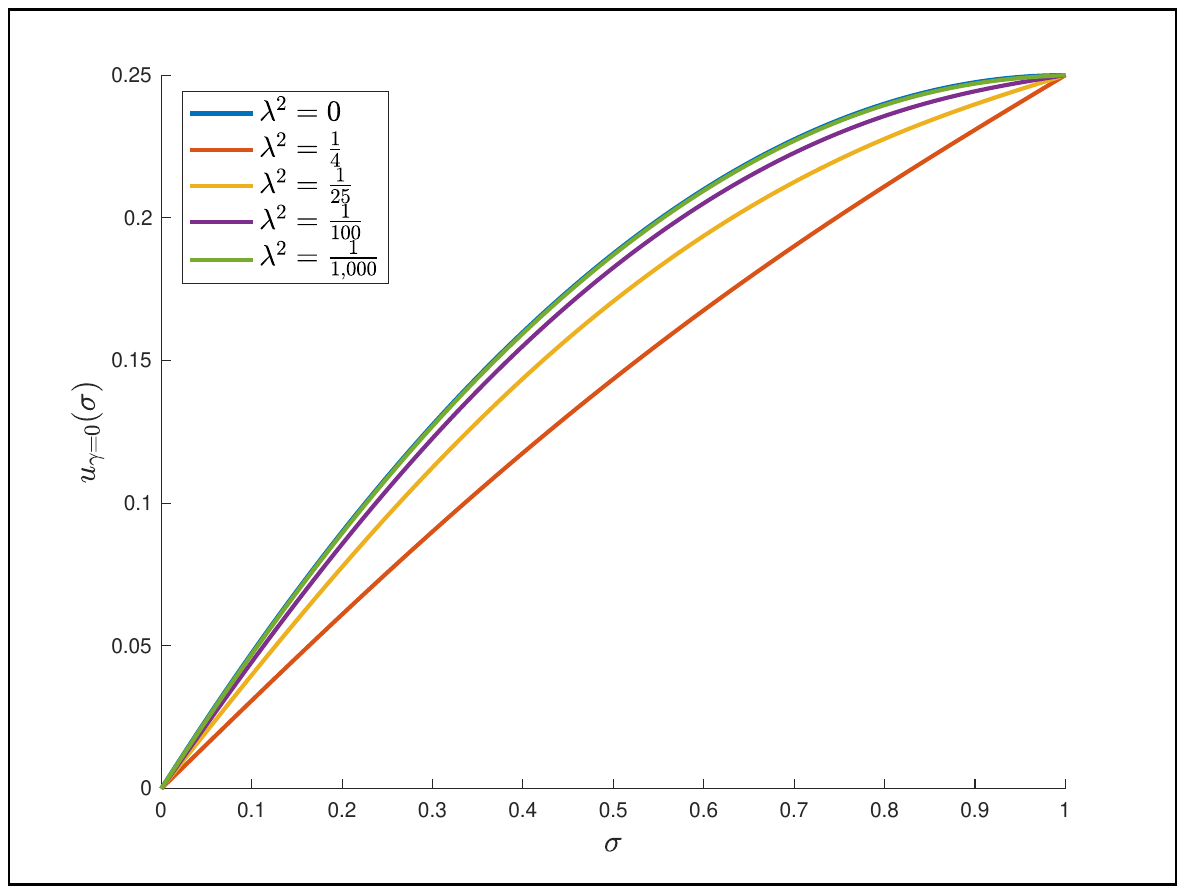}
        \caption{$\alpha = \pi/6$}
        \label{fig:f4}
    \end{subfigure}
    \hfill
    \begin{subfigure}[b]{0.49\textwidth}
        \centering
        \includegraphics[trim = {0 6cm 0 6cm},clip,width=\textwidth]{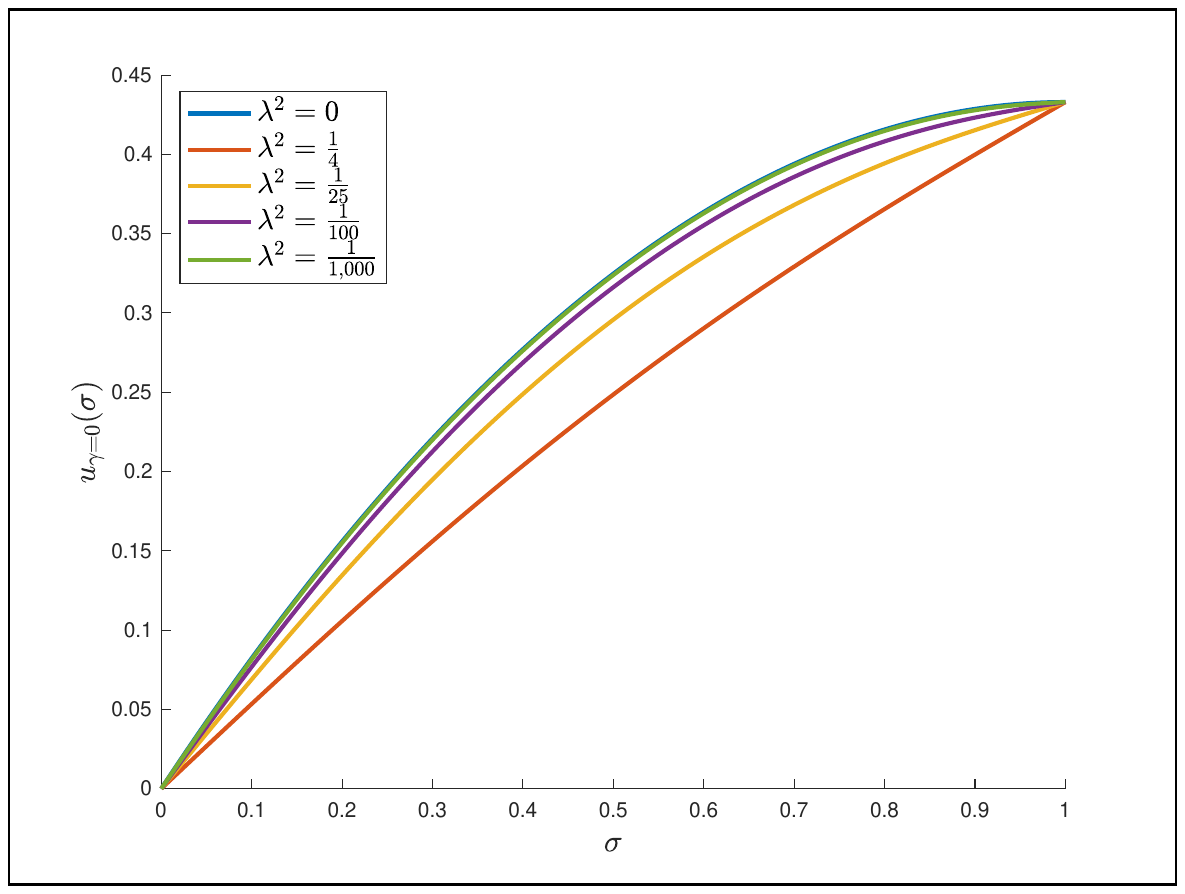}
        \caption{$\alpha = \pi/3$}
        \label{fig:f5}
    \end{subfigure}
    \caption{Graphs of the dimensionless velocity $u_{\gamma = 0}(\sigma)$ with varying values of $\lambda^2$. The curve corresponding to $\lambda^2 = 0$ represents the classical solution $u_{c, \gamma = 0}(\sigma)$.}
\end{figure}

\begin{figure}
    \centering
    \begin{subfigure}[b]{0.49\textwidth}
        \centering
        \includegraphics[trim = {0 6cm 0 6cm},clip,width=\textwidth]{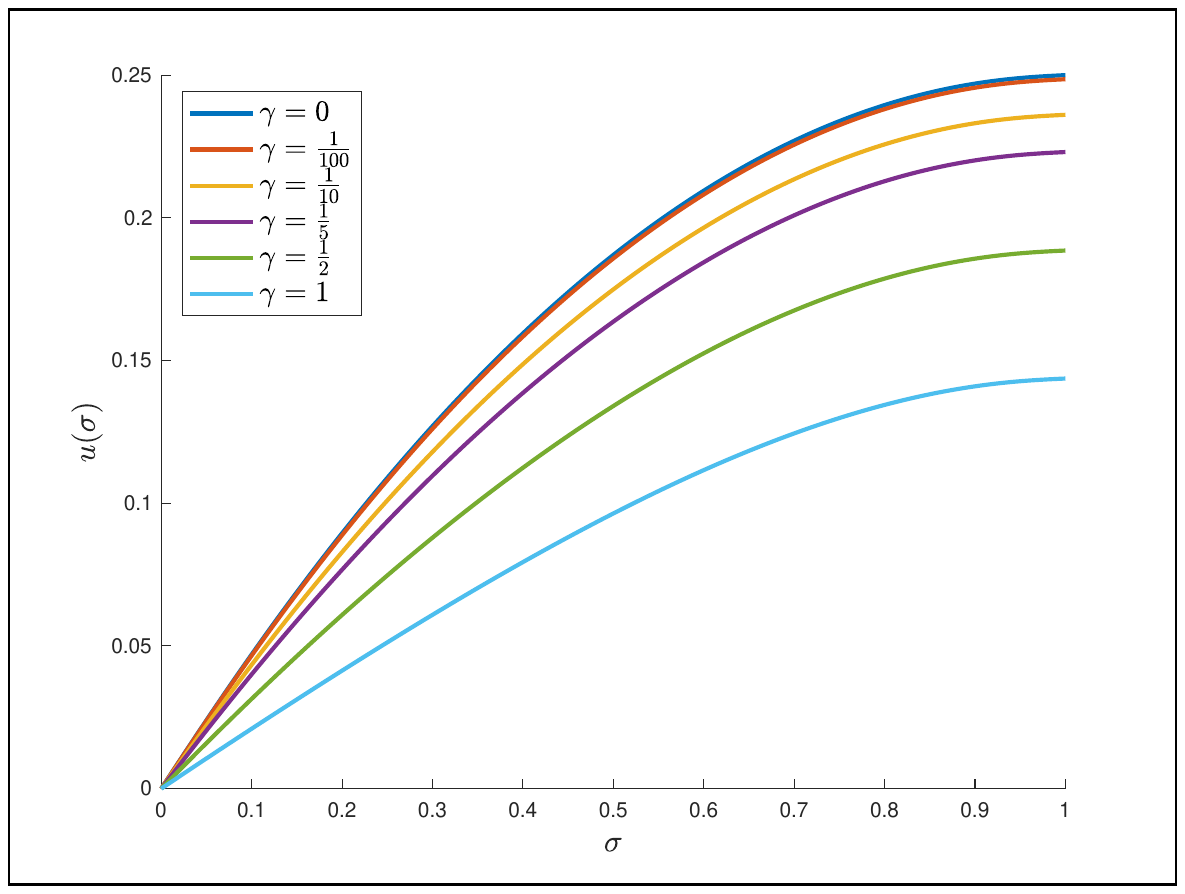}
        \caption{$\alpha = \pi/6, \pi_1 = 0, \lambda^2 = 1/1,000$}
        \label{fig:f6}
    \end{subfigure}
    \hfill
    \begin{subfigure}[b]{0.49\textwidth}
        \centering
        \includegraphics[trim = {0 6cm 0 6cm},clip,width=\textwidth]{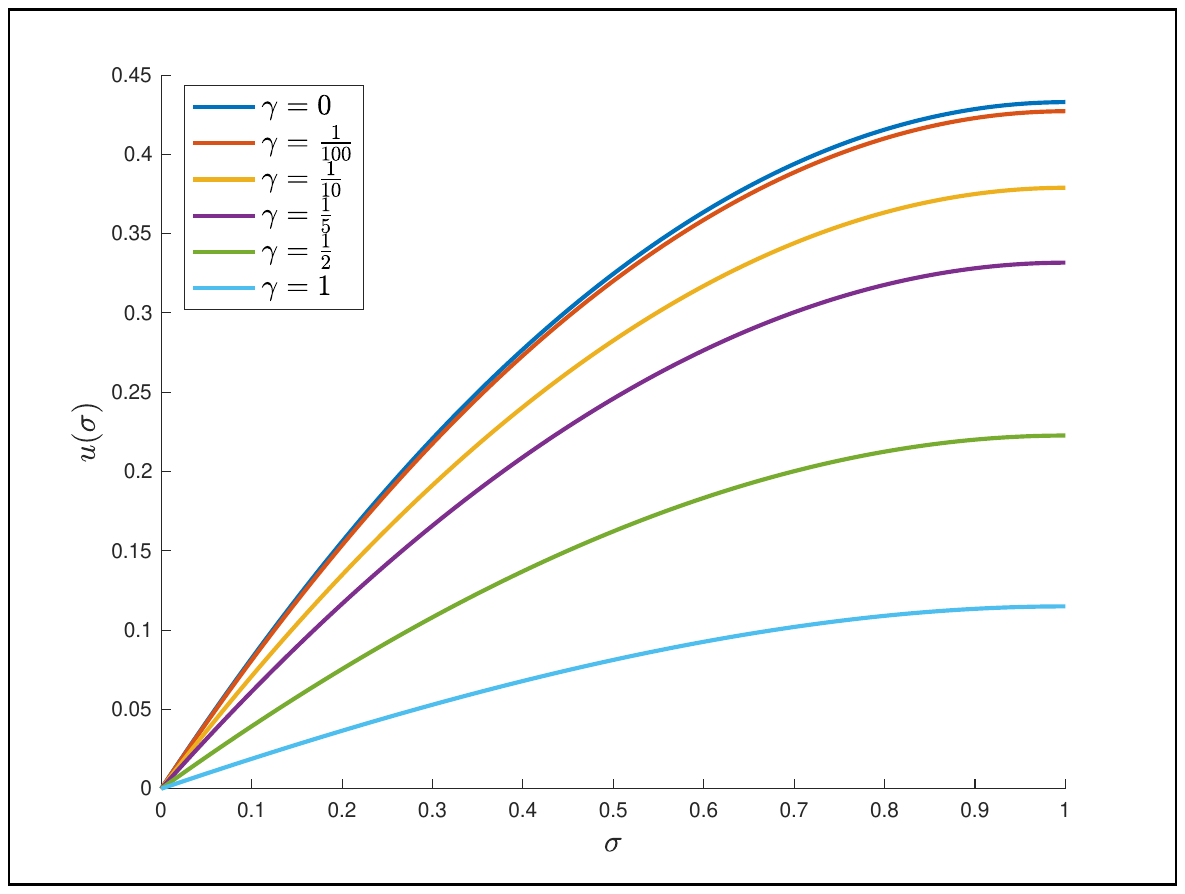}
        \caption{$\alpha = \pi/3, \pi_1 = 1, \lambda^2 = 1/10,000$}
        \label{fig:f7}
    \end{subfigure}
    \caption{Graphs of the dimensionless velocity $u(\sigma)$ with varying values of $\gamma$.}
\end{figure}

\begin{figure}
    \centering
    \begin{subfigure}[b]{0.49\textwidth}
        \centering
        \includegraphics[trim = {0 6cm 0 6cm},clip,width=\textwidth]{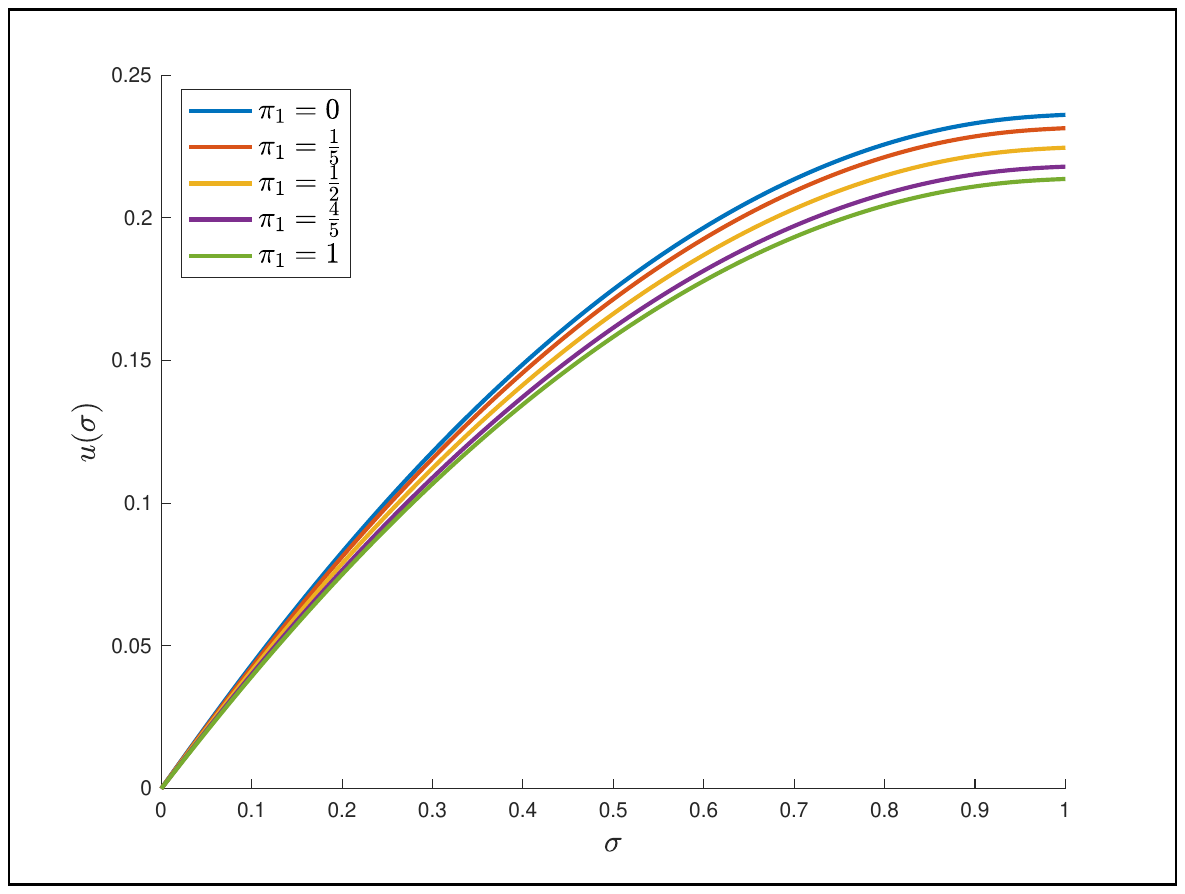}
        \caption{$\alpha = \pi/6, \gamma = 1/10, \lambda^2 = 1/1,000$}
        \label{fig:f8}
    \end{subfigure}
    \hfill
    \begin{subfigure}[b]{0.49\textwidth}
        \centering
        \includegraphics[trim = {0 6cm 0 6cm},clip,width=\textwidth]{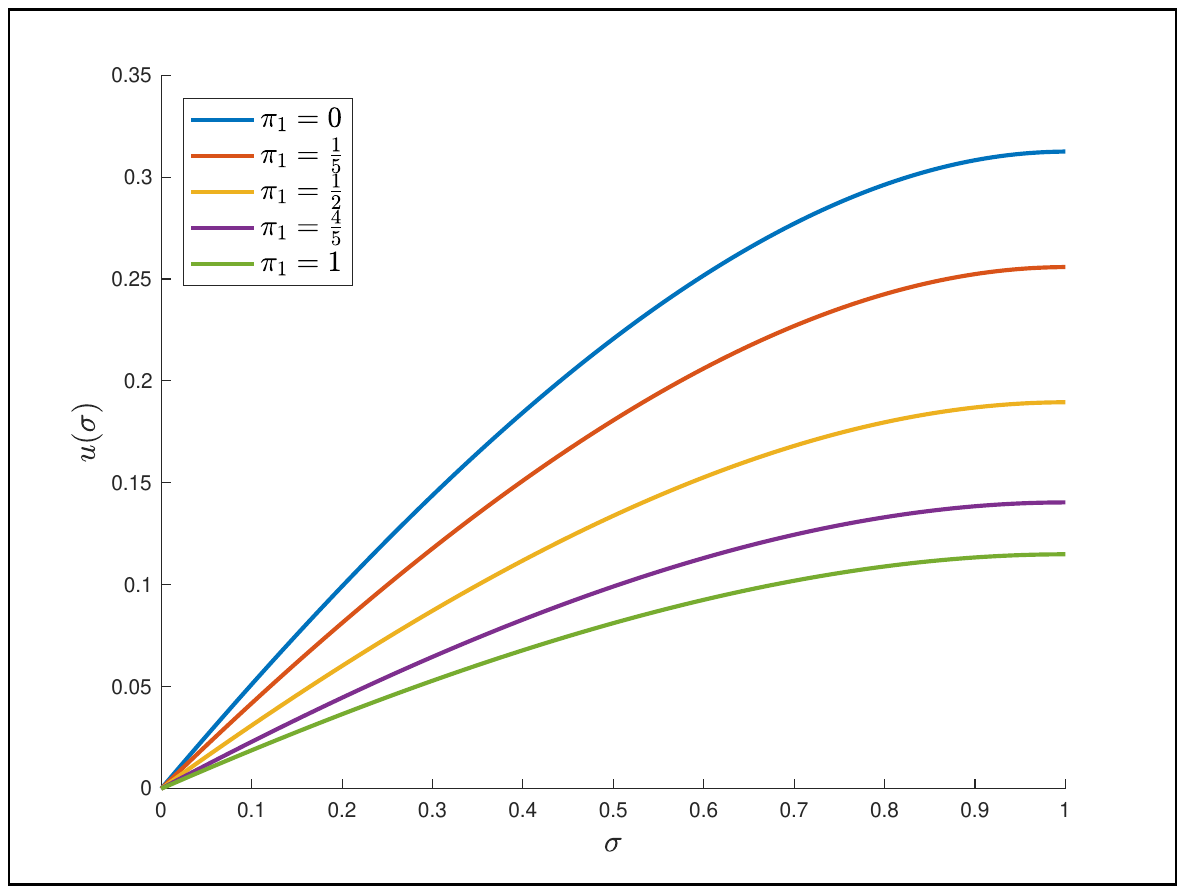}
        \caption{$\alpha = \pi/3, \gamma = 1, \lambda^2 = 1/10,000$}
        \label{fig:f9}
    \end{subfigure}
    \caption{Graphs of the dimensionless velocity $u(\sigma)$ with varying values of $\pi_1$.}
\end{figure}

\begin{figure}
    \centering
    \begin{subfigure}[b]{0.49\textwidth}
        \centering
        \includegraphics[trim = {0 6cm 0 6cm},clip,width=\textwidth]{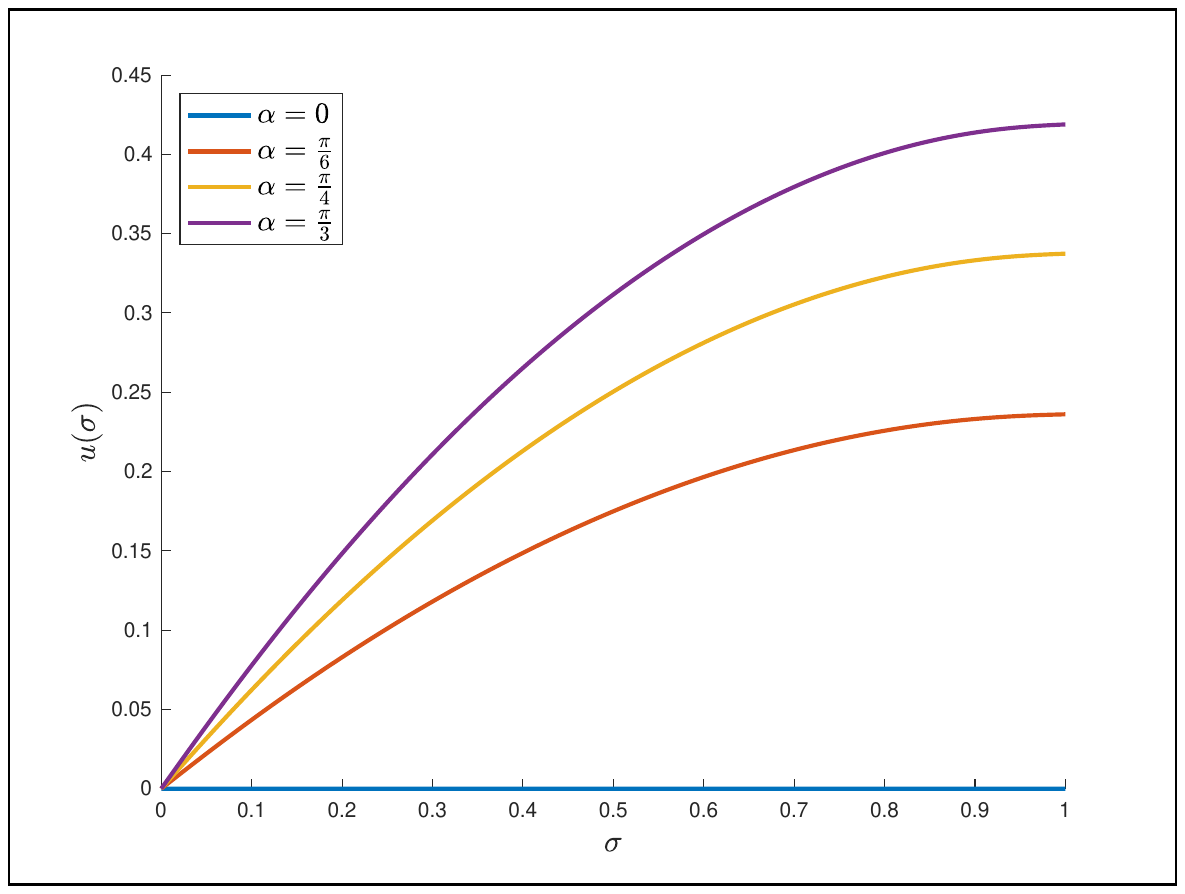}
        \caption{$\lambda^2 = 1/1,000, \pi_1 = 0, \gamma = 1/10$}
        \label{fig:f10}
    \end{subfigure}
    \hfill
    \begin{subfigure}[b]{0.49\textwidth}
        \centering
        \includegraphics[trim = {0 6cm 0 6cm},clip,width=\textwidth]{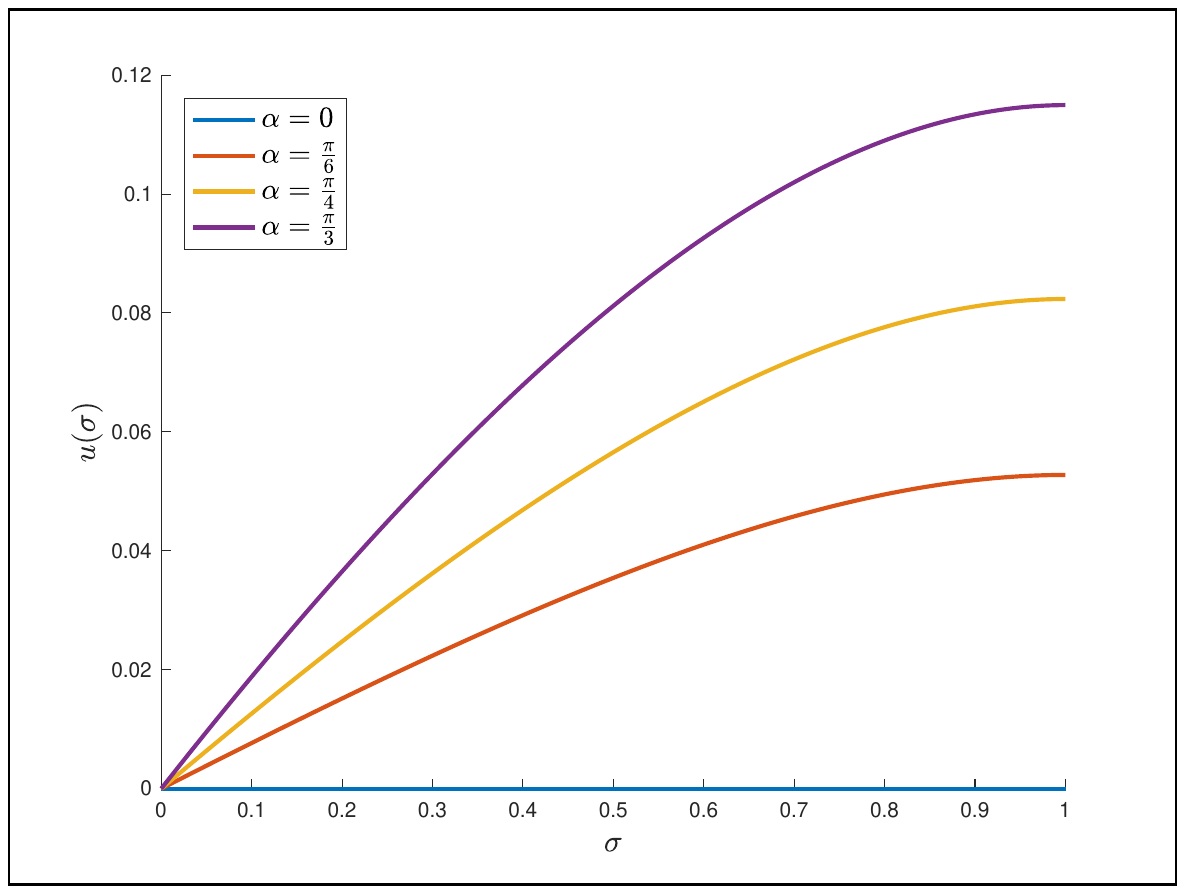}
        \caption{$\lambda^2 = 1/10,000, \pi_1 = 1, \gamma = 1$}
        \label{fig:f11}
    \end{subfigure}
    \caption{Graphs of the dimensionless velocity $u(\sigma)$ with varying values of $\alpha$. }
\end{figure}

We recall the boundary value problem governing the dimensionless velocity \eqref{eq:nondvelocitypressuredep},
\begin{equation}
	\begin{cases}
		\la^2 u'''(\sigma) - \exp(\gamma \pi(\sigma)) u'(\sigma) = -(1-\sigma)\sin\alpha, \quad \sigma \in [0,1], \\
		u(0) = 0, u''(0) = u''(1) = 0,
	\end{cases}
\end{equation}
As far as the authors are aware, this equation is not solvable explicitly, but approximate solutions can be obtained numerically. We obtain solutions of the equations with three of the parameters of interest fixed and one varying using the \textsc{bvp4c} \textsc{MATLAB} solver, see Figures \ref{fig:f2} through \ref{fig:f11}.

We now briefly discuss the physical interpretation of the solutions' behaviors. As the dimensionless length scale $\lambda$ decreases, the velocity profiles converge to the classical solution corresponding to the absence of second-gradient effects; see Figures \ref{fig:f2} and \ref{fig:f3}. The dimensionless parameter $\gamma = \beta \rho g h$ is a nondimensionalized Barus parameter and quantifies the sensitivity of viscosity to pressure. Notably, increasing the Barus parameter from $1/10$ to $1$ results in a narrower band of velocity profiles in Figure \ref{fig:f3} compared to Figure \ref{fig:f2} and Figure \ref{fig:f4}, as $\lambda$ varies. This suggests that, for larger $\gamma$, the pressure-dependent viscosity term dominates the behavior of the solutions, even when second gradient contributions are incorporated in \eqref{eq:Gpressuredependent}. In addition, an interesting observation is that when viscosity depends on pressure $(\gamma \neq 0)$, the velocity profiles obtained outspeed the classical solution in a neighborhood of $\sigma = 1$; see Figures \ref{fig:f2} and \ref{fig:f3}. This is in contrast with the constant viscosity solutions obtained in \cite{BalitactacRodriguez25} and in this article, which sit below the classical solution for all values of $\sigma$ $\in (0,1)$ and $\lambda$; see Figures \ref{fig:f4} and \ref{fig:f5}. 

Figures \ref{fig:f6} and \ref{fig:f7} show that as this sensitivity increases and the pressure dependence of viscosity becomes stronger, the flow slows down. This trend is further illustrated in Figures \ref{fig:f8} and \ref{fig:f9}, where increasing the dimensionless ambient pressure $\pi_1$ results in a substantial reduction in velocity.

Finally, as expected, increasing the incline angle $\alpha$, and thereby steepening the slope, increases the flow velocity, as shown in Figures \ref{fig:f10} and \ref{fig:f11}. In the limiting case $\alpha = 0$, corresponding to a flat surface, the solution is stationary. Collectively, these observations underscore the significant influence of both geometric parameters (such as slope angle) and constitutive features (such as pressure-dependent viscosity) in determining the flow behavior.

\section{Conclusion}
In this paper, we present the first application of the second-gradient incompressible viscous fluid model with pressure-dependent viscosity, introduced in \cite{BalitactacRodriguez25}, to the classical problem of steady flow down an inclined plane at angle $\alpha$, with weak adherence along the base and ambient pressure acting on the top free surface. We derived the simplified governing equations and boundary conditions for this setting, established well-posedness of the resulting boundary value problem, and obtained closed-form solutions in the limiting cases of constant viscosity or vanishing second-gradient effects. For the full second-gradient model with viscosity varying according to Barus' empirical relation, we computed numerical solutions and investigated how the velocity profile responds to changes in the nondimensionalized Barus parameter, ambient pressure, and slope angle. 

To date, the second-gradient models introduced in \cite{BalitactacRodriguez25} (with or without pressure dependence) have been explored only in steady-state settings. A natural and compelling research direction is to examine their behavior in time-dependent flows (e.g., oscillating flows) and to clarify in what sense they deviate from the classical Navier–Stokes theory in dynamic regimes. More broadly, a rigorous well-posedness theory for these models remains open in both the constant and pressure-dependent viscosity cases. 


\newpage 

\bibliography{researchbibmech_NSF,researchbibmech}
\bibliographystyle{plain}
\renewcommand*{\bibliofont}{\normalfont \small}

\bigskip

\centerline{\scshape C. Balitactac}
\smallskip
{\footnotesize
	\centerline{Department of Mathematics, University of North Carolina}
	
	\centerline{Chapel Hill, NC 27599, USA}
	
	\centerline{\email{corbindb@unc.edu}}
}

\bigskip

\centerline{\scshape C. Rodriguez}
\smallskip
{\footnotesize
	\centerline{Department of Mathematics, University of North Carolina}
	
	\centerline{Chapel Hill, NC 27599, USA}
	
	\centerline{\email{crodrig@email.unc.edu}}
}

\end{document}